\documentclass[12pt]{iopart}
\usepackage{graphicx}

\def\Es{{E^\star}}
\def\Eus{{{E_1}^\star}}
\def\Ezs{{{E_0}^\star}}
\def\Eeq{{{\mathcal{E}}_{\rm eq}}}

\newcommand{\sqrta}[1]{ {(#1)^{1/2}} }

\def\dsdr{\frac{\partial}{\partial r}}
\def\intzi{\int_{0}^{+\infty}}
\def\2sp{\frac{2}{\pi}}
\def\1sK{\frac{1}{\cal K}}

\begin{document}
\title{Adhesive Contact to a Coated Elastic Substrate}
\author{E. Barthel\footnote{Corresponding author:
etienne.barthel@saint-gobain.com} and A. Perriot}

\address{Surface du Verre et Interfaces, CNRS/Saint-Gobain, UMR 125, 93330, Aubervilliers Cedex France.}

\begin{abstract}
We show how the quasi-analytic method developed to solve linear
elastic contacts to coated substrates~(Perriot A. and Barthel E.
{\em J. Mat. Res.}, {\bf 2004}, {\em 19}, 600) may be extended to
adhesive contacts. Substrate inhomogeneity lifts accidental
degeneracies and highlights the general structure of the adhesive
contact theory. We explicit the variation of the contact variables
due to substrate inhomogeneity. The relation to other approaches
based on Finite Element analysis is discussed.
\end{abstract}


\section{Introduction}\label{SecIntro}
In elastic contact problems, it is known that the
homogeneous substrate is a special case which leads to
counterintuitive results. As an example, for {\it adhesionless
contacts}, the relation between penetration $\delta$ and contact
radius $a$ is independent from the {\em mechanical parameters}
(Young's modulus $E$, Poisson ratio $\nu$, reduced modulus
$\Es=E/(1-\nu^2)$) and can be in all generality expressed as a
function of the {\em shape} of the contacting bodies $z(r)$ only.

For instance, for a rigid cone of half included angle $\beta$ in
frictionless contact with an elastic substrate:
\begin{equation}\label{EqPenetrationCone}
  \delta=\frac{2}{\pi}\frac{a}{\tan\beta}
\end{equation}
Obviously, the reduced modulus is absent and one might conclude that
the relation has a geometrical origin. To show that this is not the
case, one needs only remove the assumption of homogeneous substrate.
Then, for a coated substrate, for instance, the penetration will
again depend upon the mechanical parameters of the system, in a
highly non trivial manner~\cite{Perriot04}. Overlooking the more
complex penetration dependence upon contact radius might actually
result in inaccuracies in the determination of thin film mechanical
properties by nanoindentation~\cite{Saha02,Jung04,Han06,Barthel06}.

Similarly, for adhesive contacts, a counterintuitive result is that
the {\it adhesion force} or pull-out force $F_{pull out}$ for a
sphere of radius $R$ and adhesion energy $w$ is independent from the
mechanical parameters. It is valued
\begin{equation}\label{EqFPullOut}
  F_{pull out}=n\pi w R
\end{equation}
where various $n$ values have been proposed such as
$1$~\cite{Derjaguin34}, $3/2$~\cite{JKR} or 2~\cite{DMT}.

However, changing the geometry from sphere to cone~\cite{Maugis00}
or removing the assumption of a homogeneous substrate, for example
with a coated substrate~\cite{Johnson01,Sridhar04}, again results in
reintroducing the dependence upon mechanical parameters.

With the idea that the real structure of the elastic contact models
only appears when considering a non homogeneous substrate, the
present contribution aims at: 1) showing how the recently developed
quasi analytical method for contacts to coated
substrates~\cite{Perriot04} may be used to account for adhesion as
previously used by Mary et al.~\cite{Mary07} 2) exploring how the
different concepts developed for homogeneous adhesive contacts may
be applied to a {\it non homogeneous system}, to wit, a coated
substrate.

Section~\ref{SecConcepts} reviews the various concepts which have
been developed to establish the relation between macroscopic contact
variables $F$, $\delta$ and $a$ and the physical phenomena of
adhesion. Section~\ref{SecTransform} describes the surface stress
transform method pioneered by Sneddon. Section~\ref{SecHomo}
explores the application of the method to the adhesive contact to a
homogeneous substrate, while section~\ref{SecHetero} explores the
application of the method to coated substrates.

\section{Adhesive elastic contact concepts and previous results on coated substrates}\label{SecConcepts}

\subsection{Elastic adhesive contact - problem description}

\begin{figure}
\begin{center}
\includegraphics[width=3.25in]{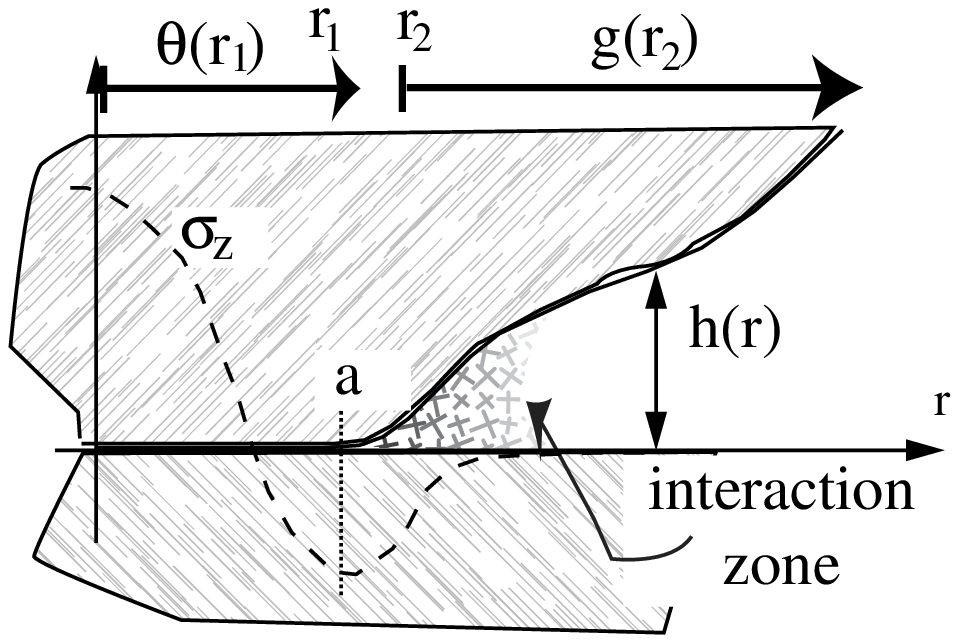}
\caption{}\label{FigGap1}
\end{center}
\end{figure}

Contact problems are characterized by mixed boundary conditions: the
surface displacement is specified inside the contact zone and the
normal stress at the surface outside. Note that this normal stress
outside the contact zone is zero, except at the edge of the contact
zone where more or less long range adhesive surface forces develop
in the so-called interaction zone (Fig.~\ref{FigGap1}).

Three concepts have been developed to handle these complex boundary
conditions, provide a solution and establish relations between the
macroscopic variables $F$, $\delta$ and $a$.

\subsection{Three concepts to solve the adhesive elastic contact
problem}\label{SecConcepts2}

\subsubsection{Energy minimization}\label{SectionMacro} The first method disregards the contact
edge details, assumes a given form for the surface stresses inside
the contact zone and calculates the solution parameters by energy
minimization, taking into account the adhesion energy $w$ as a $\pi
w a^2$ term. This macroscopic approach was implemented as early as
1934, when Derjaguin assumed a Hertz-like stress distribution and
combined it with an adhesion energy contribution proportional to the
contact area, which resulted~\cite{Derjaguin34} in a pull-out force
given by $n=1$ in Eq.~\ref{EqFPullOut}.

Based on elastomer adhesion experiments, a more realistic surface
deformation built up from the addition of a flat punch displacement
to the Hertz solution resulted in the JKR
model~\cite{JKR}~\footnote{Note that an early version of this model
can be found in the appendix of G. Sperling's PhD
thesis~\cite{Sperling64}, where it stands as a curiosity, not as the
"Theorie der Haftung" developped in the body of the text, which
mainly deals with surface roughness and plastic deformation.} where
$n=3/2$ in Eq.~\ref {EqFPullOut}. This model, now generally accepted
for soft solids, provides a more consistent description of the
stresses inside the contact zone than the original 1934 result.

More generally, a formal method to minimize the total energy with
regards to the contact radius $a$ naturally leads to introducing the
energy release rate, a concept borrowed from fracture mechanics : if
$\cal E$ is the stored elastic energy, the energy release rate $G$
can be defined in an axisymmetric geometry as
\begin{equation}\label{EqDefEnergyReleaseRate}
 2\pi a G(a)= \left.\frac{\partial{\cal E}}{\partial a}\right|_\delta
\end{equation}
Such an approach was used for instance as an alternative derivation
of the JKR model in~\cite{Johnson85}.


\subsubsection{Stress intensity factor}\label{SectionFractureMechanics}
The second method is to focus on the crack-like stress singularity
at the contact edge. It is described in terms of the {\em stress
intensity factor}. The relation between the energy release rate and
the stress intensity factor was initially proposed by
Irwin~\cite{Irwin57}. In the field of the adhesive contact, this
second method was mainly propounded by
Maugis~\cite{Maugis00,Maugis78,Maugis87}. It is particularly useful
for direct extensions of the Linear Elastic Fracture Mechanics where
a description in terms of stresses at the crack tip is necessary:
for instance, it may be used to include crack tip viscoelasticity
~\cite{Maugis78,Maugis87,Greenwood81}.

\subsubsection{Exact solution}\label{SectionExact}
The third method is actually more general: it takes into account the
details of the attractive stress distribution outside the contact
zone. This idea was somehow initiated by the DMT model~\cite{DMT}
($n=2$ in Eq.~\ref{EqFPullOut}) but was brought to a new dimension
by the introduction of a cohesive zone model by
Maugis~\cite{Maugis92}, which resulted in a clear exposition of the
transition between the two limiting models JKR and DMT. It also led
to the development of the fully viscoelastic adhesive contact
models~\cite{Barthel02,Haiat03}. In these models, stress relaxation
inside the contact zone -- as in the adhesionless viscoelastic
contact by Ting~\cite{Ting66} -- and creep inside the interaction
zone -- as in the viscoelastic crack
models~\cite{Maugis78,Maugis87,Greenwood81} -- are taken into
account simultaneously.

Just as the viscoelastic case extends the  adhesive contact model
from instantaneous response to time dependent constitutive relation,
the coated substrate generalizes the model to an inhomogeneous half
space. Our interest here will however be limited to the small
cohesive zone sizes, located on the JKR side of the transition,
which connect this approach to the previous two. For an extension to
the JKR-DMT transition with a coated substrate, when the interaction
zone grows to extensions comparable or larger than the film
thickness, in the spirit of the Maugis model, see~\cite{Sergici06}.

\subsection{Review of previous results on the adhesive contact to coated substrates}
On the theoretical side, the literature on the adhesive contact to
coated substrates is scanty, although, in practice, thin soft
adhesive layers are often used, as in many applications of pressure
sensitive adhesives and functional or protective organic coatings.
Actually only two series of papers by Shull and
coworkers~\cite{Shull98,Shull02} on one side and Johnson and
Sridhar~\cite{Johnson01,Sridhar04} on the other side have been
published.

Shull and coworkers calculate the energy release rate from the
contact compliance (cf.~\ref{SectionCompliance}). To complete the
calculations, the compliance of coated sustrates are computed by
Finite Element (FE) methods and phenomenological fits or numerical
differentiation may be carried out.

Johnson and Sridhar's approach goes along the line of the stress
intensity factor (section~\ref{SectionFractureMechanics}) approach:
still using FE, they directly calculate the energy release rate $G$
by a stiffness matrix variation method~\cite{Parks74} and extract
stress intensity factors from $G$. From this stress intensity
factor, they build the adhesive contact relations.

Both methods converge on their ultimate use of Finite Element
calculations to handle the complexities of the coated substrate
response. Appearances to the contrary, these two series of papers
seem actually to rely on the same method and an in depth
investigation of the Parks method in the context of the frictionless
contact would probably connect the stiffness matrix derivative used
by Parks to the contact stiffness derivative method as developped by
Shull.

 In the present paper, we show that the quasi-analytical method introduced previously~\cite{Perriot04} may be used
 to efficiently compute all the necessary contact variables and in particular provide
 direct numerical evaluations for the stress intensity factor for coated
 substrates. It is also useful to expose the equivalence of the various
 concepts for the adhesive contact in the wider context of
 non-homogeneous substrates.

\section{Sneddon and the surface stress transform}\label{SecTransform}
Adequate transforms facilitate the investigation of linear elastic
axisymmetric contacts formulated as integral
equations~\cite{Gladwell80}. It is ideally suited to the adhesive
contacts because of the relevant boundary conditions including the
interaction zone.

Following our previous
studies~\cite{Barthel02,Haiat03,Barthel98,Huguet00}, we use $g(r)$,
which is a transform of the normal stress distribution at the
surface $\sigma_z$ defined as
\begin{equation}\label{EqDefG}
g(r)=\int_{r}^{+\infty} \frac{s \sigma_z(s)}{\sqrt{s^2-r^2}}ds
\end{equation}
and $\theta(r)$ which is a transform of the normal surface
displacement $u_z$ defined by
\begin{equation}\label{EqDefTheta}
\theta(r)=\dsdr \int_0^r \frac{s u_z(s)}{\sqrt{r^2-s^2}}ds
\end{equation}

These relations are suited to the adhesive contact problem because
$g(r)$ is expressed as a function of normal surface stresses at
radii values larger than $r$ and $\theta(r)$ as a function of
surface normal displacement at radii values smaller than $r$, in
agreement with the adhesive contact boundary conditions
(Fig.~\ref{FigGap1}). In particular $g$ is zero outside the
interaction zone.

Similarly, inside the contact zone ($r\le a$), $u_z(r)=\delta-z(r)$
where $z(r)$ is the shape of the indenter. Integration by parts
transforms Eq.~\ref{EqDefTheta} into
\begin{equation}
  \theta(r) = \delta-\delta_0(r) \label{EqTheta}.
\end{equation}
where $\delta$ is the penetration and $\delta_0$ depends only upon
the shape of the indenter $z(r)$ through
\begin{equation}\label{EqDelta0}
\delta_0(r)= r \int_0^r \frac{z'(s) ds}{\sqrt{r^2-s^2}}
\end{equation}

The full power of these transforms appears when it is
recognized~\cite{Barthel02,Haiat03} that in the linear elastic case
and for a {\em homogeneous} substrate, mechanical equilibrium
results in
\begin{equation}\label{Equilibrium}
  g(r)=\frac{E^\star}{2}\theta(r)
\end{equation}

This provides the direct solution to the adhesive contact problem
which we detail now.

\section{Adhesive contacts on homogeneous substrates}\label{SecHomo}

Before tackling the problem of the adhesion to coated substrates we
review our present understanding of elastic contacts to homogeneous
substrates using the transform method.

\subsection{Non adhesive contact of smooth indenters}

In the absence of adhesion, Eq.~\ref{EqDefG} shows that $g(a)=0$. If
the indenter shape $z(r)$ is smooth at the contact edge, then
Eqs.~\ref{EqTheta} and \ref{Equilibrium} imply $\theta(a)=0$ and
\begin{equation}\label{EqPenetrationNonAdh}
  \delta=\delta_0(a)
\end{equation}

Thus the indenter penetration in an adhesionless contact is the
function $\delta_0(a)$, which depends only upon the geometry of the
indenter (Eq.~\ref{EqDelta0}) and not on the elastic parameters.
This is the counterintuitive result stated in
section~\ref{SecIntro}.

\subsection{Non adhesive contact of flat
punches}\label{SectionFlatPunch}

A special case however is the flat punch for which $\delta_0(r)=0$
for $r<a$ and $g(r)=0$ for $r>a$. The penetration is independent of
the contact radius as expected since the contact radius is constant,
equal to the punch radius.

Then $\theta(a^-)=\delta$ so that $g(a^+)=0$ and $g(a^-)\neq 0$. The
exact meaning of this discontinuity in the $g$ transform at $a$ will
be discussed below (section~\ref{SectionSIF}).

Note also that the force can be obtained through the simple but
general expression~\cite{Huguet00}
\begin{equation}\label{EqForceGeneral}
  F=4\int_0^{+\infty}g(s)ds
\end{equation}
Specifically here,
\begin{equation}\label{EqForceNonAdh}
  F_0(a)=4\int_0^{a}g(s)ds
\end{equation}
For the flat punch the force is then directly calculated from
Eqs.~\ref{EqTheta}, \ref{Equilibrium} and \ref{EqForceGeneral} as
\begin{equation}
  F_0(s) = S(a)\delta_{fp}
\end{equation}
with
\begin{equation}
S(a)=2aE^\star.
\end{equation}

\subsection{Adhesive contacts}

In the adhesive case, however, although the indenter shape is smooth
at the contact edges, the condition $g(a)=0$ is relaxed due to the
adhesive interactions. The core of the JKR approximation is to
neglect the details of the stresses and deformations at the contact
edge and encapsulate the adhesive contribution in an additional flat
punch displacement. This amounts to a description where the
attractive interaction stresses drop abruptly to zero outside the
contact zone, leading to a stress singularity akin to those met in
fracture mechanics.

Then Eqs.~\ref{EqPenetrationNonAdh} and \ref{EqForceNonAdh} become
\begin{eqnarray}
 \delta &=& \delta_0 + \delta_{fp}\label{EqForceHomo}\\
F &=& F_0 + S(a)\delta_{fp}\label{EqPenetrationHomo}
\end{eqnarray}

The question is to provide a relation between this additional flat
punch displacement $\delta_{fp}$ and the adhesion energy $w$ for a
given contact radius $a$.

\subsection{Various concepts for the derivation of adhesive contact results}
For future reference, we now apply the surface stress transform
solution to the three methods developed to handle adhesive contacts
(section~\ref{SecConcepts2}).

\subsubsection{Energy Release Rate}\label{SectionERRHomo} A very compact derivation is obtained when it is recognized that for a homogeneous substrate $g$ et $\theta$ are independent of $a$ inside the
contact zone. In addition the total mechanical energy $\cal E$ as a
function of the normal surface stress distribution is
\cite{Huguet00}
\begin{equation}\label{EqEnergieHomo}
{\cal E} = \frac{4}{\Es} \intzi ds~g^2(s)=\Es \intzi ds~\theta^2(s)
\end{equation}
Then for constant penetration, with $g(a)=0$ outside the contact
zone,
\begin{equation}\label{EqSelfConHomo2}
2\pi a G(a)=\frac{4g(a)^2}{\Es}=\Es\theta(a)^2
\end{equation}

Equating $G$ and the adhesion energy $w$ results in
\begin{equation}
  w=\frac{2g(a)^2}{\pi E^\star a} \label{Eqselfconsistence}
\end{equation}

The energy release rate $G(a)$ can be expressed as a function of the
local stress distribution at the crack tip because of
Eq.~\ref{EqDefG} and $g(r)=0$ for $r>a$.

\subsubsection{Stress intensity factor}\label{SectionSIF}
This expression together with the plain strain Irwin
relation~\cite{Maugis00,Irwin57} between stress intensity factor and
energy release rate\footnote{The $1/2$ factor comes from the
rigidity of the punch.}
\begin{equation}
G={K_I}^2/(2E^\star)
\end{equation}
suggests that $2g(a)/\sqrt{\pi a}$ assumes the role of a stress
intensity factor.

A direct derivation is obtained~\cite{Maugis00} if we use the
following expression for the stress intensity factor at the contact
edge
\begin{equation}
  K_I = \lim_{\epsilon\rightarrow 0} \sqrta{2\pi\epsilon} \sigma(a-\epsilon)
\end{equation}
Eq.~\ref{EqDefG} can be inverted~\cite{Huguet00} providing
\begin{equation} \label{EqSigmaG}
\sigma_z(s)=\2sp\left[ \frac{g(a) \Theta(a-s)}{\sqrta{a^2-s^2}} +
\int_s^{+\infty} dt \frac{ g'(t)}{\sqrta{t^2-s^2}} \right]
\end{equation}
where it has been assumed that $g$ is smooth except for a
discontinuity at $a$. We use the notation $g(a^-)=g(a)$ and assume
$g(a^+)=0$. Then
\begin{equation}\label{EqSIF}
   K_I = \frac{2g(a)}{\sqrt{\pi a}}
\end{equation}
which confirms that $g(a)$ has the form and meaning of a stress
intensity factor.

As a stress intensity factor, $g(a)$ is a measure of stress
singularity at the contact edge. An ancillary property of the
transform defined by Eq.~\ref{EqDefG} is that it regularizes the
singular crack like stress distributions and Eq.~\ref{EqDefTheta}
establishes a relation between the additional flat punch penetration
and the stress singularity through
\begin{equation}\label{EqThetaFP}
\theta(a)=\delta_{fp}=\frac{2 g(a)}{E^{\star}}
\end{equation}

\subsubsection{Self-consistent method}

The self-consistent description of the interaction zone
\cite{Maugis92,Barthel98,Greenwood98} can be explored at the limit
of negligible interaction zone extension, {\it i.e.} in the JKR
limit.

One possible starting point of the self-consistent
method~\cite{Barthel98,Huguet00} is to calculate the adhesion energy
by
\begin{equation}\label{EqSelfCon}
  w=-\int_a^{+\infty}\sigma(s)\frac{dh}{ds}ds
\end{equation}
Following~\cite{Haiat03}(Eq~15) the gap between the surfaces $h$ is
split into the contributions of the contact stresses and the
interaction stresses
\begin{equation}\label{EqGap}
  h(r)=h_{Hertz}(r,a)+ h_{int}(r,a)
\end{equation}
If the interaction range is small (which results in $c-a << a$), the
radial extension of the interaction will also be small, $g'$ is
peaked around $a$, and the dominant term in Eq.~\ref{EqSelfCon} will
be the $\partial u_{z,int}/\partial s$ term. Thus \cite{Huguet00},
\begin{equation}\label{EqSelfConHomo}
w\simeq-\int_a^{c}\sigma_s\frac{dh_{int}}{ds}ds\simeq -\frac{4}{\pi
a \Es}\int_a^{+\infty} g'(t)g(t)dt
\end{equation}
As a result, one again recovers Eq.~\ref{Eqselfconsistence}.
Combined with Eq.~\ref{EqSIF}, this method may be viewed as a direct
derivation of the Irwin relation~\cite{Irwin57}.

\subsection{Normalized form}
The adhesive contact equations are obtained by combining
Eqs.~\ref{EqForceHomo}-\ref{EqPenetrationHomo} and
Eq.~\ref{Eqselfconsistence}. For later reference, we specialize the
results to the spherical indenter and use the Maugis
normalization~\cite{Maugis92}
\begin{eqnarray}
  P&=&\frac{F}{\pi w R}\\
D&=&\frac{\delta}{{\left(\frac{\pi^2 w^2R}{\Es^2}\right)}^{1/3}}\\
  \bar a&=&\frac{a}{\left(\frac{\pi w R^2}{\Es}\right)^{1/3}}
\end{eqnarray}
One obtains
\begin{eqnarray}
  P = \frac{4}{3}{\bar a}^3 - 2\sqrt{2}{\bar a}^{3/2}\label{EqForceJKR} \\
  D = {\bar a}^2 - \sqrt{2}{\bar a}^{1/2}\label{EqPenetrationJKR}
\end{eqnarray}
These are the usual JKR equations which will be generalized in the
next section. Their structure directly reflects the general
structure of Eqs.~\ref{EqForceHomo}-\ref{EqPenetrationHomo} with the
identifications $\bar \delta_{fp} = -\sqrt{2}{\bar a}^{1/2}$ and
${\bar S}_{fp} = 2{\bar a}$.

\section{Coated Substrates -- Thin Films}\label{SecHetero}

\subsection{Description of the contact}
If the substrate is not homogeneous the simple diagonal equilibrium
relation Eq.~\ref{Equilibrium} between $g$ and $\theta$ is lost.
However, keeping the same transforms, we have shown~\cite{Perriot04}
that a useful relation subsists
\begin{equation}\label{EqEquilibriumCS}
  \theta(r;a,t,[E])=\frac{2g(r;a,t,[E])}{\Eus}+\2sp\int_0^{+\infty}g(s;a,t,[E])K(r,s;t,[E]) ds
\end{equation}
where $K$ expresses the elastic response of the coated
substrate~\cite{Perriot04}, $t$ stands for the coating thickness and
$[E]$ for the four mechanical parameters (film and substrate Young's
moduli and Poisson ratios). Since $g$ is zero outside the contact
zone, the upper boundary will typically be the contact radius $a$ or
the interaction zone radius $c$ if $c\neq a$. The finite integration
interval facilitates the numerical inversion~\cite{Perriot04} of
Eq.~\ref{EqEquilibriumCS}. Note that direct analytical inversion is
impossible because of the complex expression for $K$.

We now split the stress distribution and the displacement field into
their non adhesive $\delta_H$ and flat punch $\delta_{fp}$
components.

For the non adhesive contact of an indenter of regular shape, the
penetration is given by
\begin{eqnarray}
  \delta_H(a,t,[E])-\delta_0(r)&=&\frac{2g_H(r;a,t,[E])}{\Eus} \\
  &+&\2sp\int_0^a g_H(s;a,t,[E])K(r,s;t,[E])ds \label{EqPenetrationInhom}
\end{eqnarray}
where $\delta_0(r)$ reflects the indenter shape and $g_H(a;a,t,[E])
= 0$ in the absence of adhesion, ensuring a unique solution.
Eq.~\ref{EqPenetrationInhom} generalizes Eq.~\ref{EqPenetrationHomo}
and introduces the non-trivial relation mentioned in
section~\ref{SecIntro}.

For the contact of the {\it flat punch} the problem is linear with
$\delta_{fp}$ and
\begin{equation}\label{EqEquilibriumFPInhom}
  \delta_{fp}=\frac{2g_{fp}(r;a,t,[E])}{\Eus}+\2sp\int_0^ag_{fp}(s;a,t,[E])K(r,s;t,[E])
  ds
\end{equation}
with $g_{fp}(r;a,t,[E])\propto\delta_{fp}$ and $g(a;a,t,[E])\neq0$.
Eq.~\ref{EqEquilibriumFPInhom} generalizes Eq.~\ref{EqThetaFP} into
a non trivial proportionality relation between $g(a)$ and the flat
punch penetration $\delta_{fp}$.

Then the penetration for the adhesive contact of an indenter of
arbitrary shape is
\begin{equation}\label{EqPenetrationCS}
\delta(a,t,[E])=\delta_H(a,t,[E])+\delta_{fp}
\end{equation}
With the following definition of the contact stiffness
\begin{equation}\label{EqStiffnessInhom}
S(a,t,[E])=\frac{F_{fp}(a,t,[E])}{\delta_{fp}}
\end{equation}
the force is
\begin{equation}\label{EqForceCS}
F(a,t,[E])=F_H(a,t,[E])+ S(a,t,[E])\delta_{fp}
\end{equation}

Equations~\ref{EqPenetrationCS} and \ref{EqForceCS} generalize
Eqs.~\ref{EqPenetrationHomo} and \ref{EqForceHomo}. Here again, the
question is to provide a relation between the additional flat punch
displacement $\delta_{fp}$ and the adhesive interaction.

\subsection{Derivation}

For coated substrates, the resolution benefits from more general
expressions derived for viscoelastic adhesive
contacts~\cite{Barthel04}. This is not fortuitous but results from a
similar breakdown of the simplified relations derived through
Eq.~\ref{Equilibrium} when spatial heterogeneity or time dependence
is introduced.

\subsubsection{Energy Release Rate}

The total energy is~\cite{Barthel04}
\begin{equation}\label{EqEnergieHetero}
{\cal E} = 2 \intzi ds~g(s)\theta(s)
\end{equation}
Then
\begin{equation}\label{EqGHetero} 2\pi aG(a)=
2\left(g(a)\theta(a)+ \int_0^a ds~\frac{dg(s)}{da}\theta(s)\right)
\end{equation}
because $G(a)$ is calculated at constant displacement so that inside
the contact zone, the surface displacement -- and therefore
$\theta(r)$ -- are unaffected by the additional stress distribution
$g(a)$: there is only a stress rearrangement inside the contact zone
which does affect $g(r), r<a$ for a coated substrate: indeed the
local relation Eq.~\ref{Equilibrium} between $g$ and $\theta$ breaks
down in this case.

Following Mary et al.~\cite{Mary07}, one can show that this non
local contribution is actually canceled by the non local
contribution included in the $g(a)\theta(a)$ term. For $s<a$
\begin{equation}
  -\frac{\pi}{\Eus}\frac{dg(s)}{da}=g(a)K(a,s)+\int_0^a
  dr~\frac{dg(r)}{da}K(r,s)
\end{equation}
and multiplying by $g(s)$ and integrating between 0 and $a$, one
obtains
\begin{equation}
g(a)\theta(a)+ \int_0^a
dr~\frac{dg(r)}{da}\theta(r)=\frac{2g(a)^2}{\Eus}
\end{equation}

The region affected by the crack tip stresses extend over a distance
commensurate with the interaction zone size, {\it i.e.} it is
smaller than the film thickness so that the energy release rate is
controlled by the {\it film} compliance.



\subsubsection{Compliance Method}\label{SectionCompliance}
The compliance formulation is at the core of the method used by
Shull and coworkers but the derivation given in some of their
earlier papers was obscured by unnecessary assumptions. This
formulation emerges readily when $\cal E$ in
Eq.~\ref{EqEnergieHetero} is expressed as a function of strain
instead of stress\footnote{Such a {\it strain} energy release rate
has also been introduced previously in viscoelastic crack problems
by Schapery~\cite{Schapery89} and Greenwood and
Johnson~\cite{Greenwood81}.}.

We split the total stress and strain fields in their non adhesive
($g_H$ and $\theta_H$) and flat punch components ($g_{fp}$ and
$\theta_{fp}$). This allows for easy integration because
$\theta_{fp}(r;a,t,[E])=\delta_{fp}$ is independent of $r$. Using
Betti's theorem to calculate the cross terms
\begin{eqnarray}
&&2 \intzi ds g_H(s;a,t,[E])\theta_{fp}(s;a,t,[E]) \\
&=& 2 \intzi ds g_{fp}(s;a,t,[E])\theta_H(s;a,t,[E])\\
&=&\frac{1}{2}\delta_{fp}P_H(a,t,[E])
\end{eqnarray}
and similarly using Eq.~\ref{EqStiffnessInhom}
\begin{equation}
2 \intzi ds
g_{fp}(s;a,t,[E])\theta_{fp}(s;a,t,[E])=\frac{1}{2}S(a,t,[E])\delta_{fp}^2
\end{equation}
to express the flat punch elastic energy, one obtains the total
energy
\begin{equation}
  {\cal E}(a,
  \delta_{fp})=U_H(a,t,[E])+\frac{1}{2}S(a,t,[E])\delta_{fp}^2+P_H(a,t,[E])\delta_{fp}
\end{equation}
where $\delta_fp$ is negative.

\begin{figure}
\begin{center}
\includegraphics[width=3.25in]{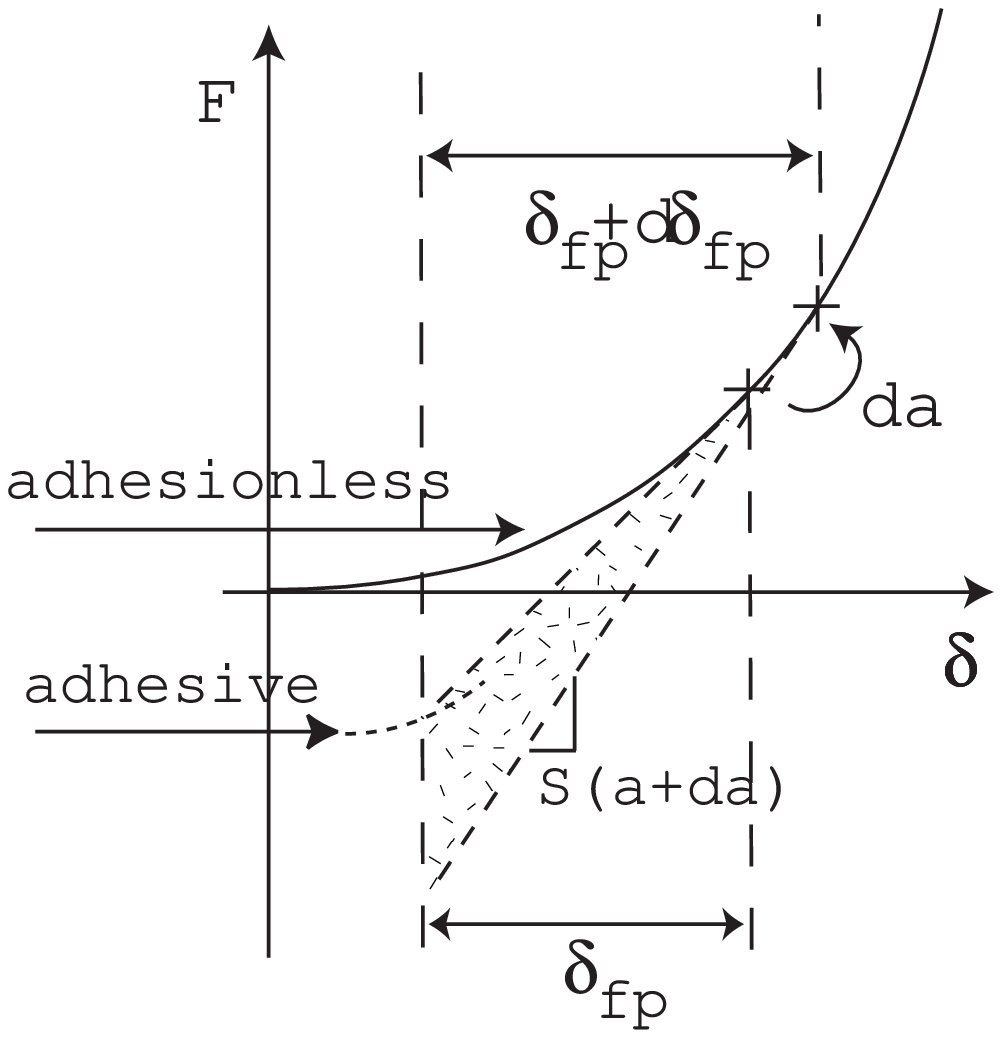}
\caption{}\label{FigGgraph}
\end{center}
\end{figure}

A graphic illustration of this result is given in
Fig.~\ref{FigGgraph}.

The energy release rate $G$ is the differential of the total energy
${\cal E}(a,\delta_{fp})$ with respect to contact area
(Eq.~\ref{EqDefEnergyReleaseRate}) at constant total penetration
$\delta$ where
\begin{equation}
  \delta = \delta_H(a) + \delta_{fp}
\end{equation}

\begin{equation}
\left.\frac{d{\cal E}}{da}\right|_\delta = \frac{\partial {\cal
E}}{\partial a}+\frac{\partial {\cal E}}{\partial
\delta_{fp}}\frac{d\delta_{fp}}{da}
\end{equation}
Now
\begin{eqnarray}
  dU_H&=&P_H(a,t,[E]) d\delta_H\\
  dP_H&=&S(a,t,[E])d\delta_H\label{EqRaideur}
\end{eqnarray} so that finally all terms cancel\footnote{Due to the counterintuitive results obtained for the homogeneous substrate,
one may develop doubts about the identity of the stiffness defined
by Eq.~\ref{EqRaideur} and the stiffness defined by
Eq.~\ref{EqStiffnessInhom}. That this identity does hold for a non
homogeneous substrate is shown in the appendix
(section~\ref{SecAppendix}. It is due to the absence of adhesion:
the normal stress at the edge of the contact is zero so that contact
radius variation does not result in force variation, to first
order.} except the differential of the flat punch elastic energy and
\begin{equation}\label{EqCompliance}
2\pi a G(a) = \frac{1}{2}\delta_{fp}^2\frac{dS}{da}
\end{equation}

This differentiation process is also illustrated on
Fig.~\ref{FigGgraph}.

\subsubsection{Stress Intensity Factor}
The derivation of the stress intensity factor expression in
section~\ref{SectionSIF} is indendent of the material properties and
only results from the definition of the transform $g$. It is
therefore unchanged in the more general case and the pending problem
is actually the relation between $g(a)$ and the energy release rate
$G$.

\subsubsection{Self-consistency}
In Eq.~\ref{EqGap},
\begin{equation}
h_{int}(r,a,t,[E])=\2sp\int_a^r\frac{\theta(s)-\theta(a)}{\sqrt{r^2-s^2}}ds
\end{equation}
If at the crack tip the deformation predominantly results from the
adhesive interactions -- and this is the essence of the JKR limit --
then Eq.~\ref{EqSelfCon} becomes~\cite{Barthel02}
\begin{equation}
  w=-\frac{2}{\pi a}\int_a^{c}\theta'(r)g(r)dr
\end{equation}

From Eq.~\ref{EqEquilibriumCS} we obtain
\begin{equation}\label{EqEquilibriumDeriveeCS}
  \theta'(r)=\frac{2g'(r)}{\Eus}+\2sp\int_0^{+\infty}g(s)\frac{dK(r,s)}{dr}
\end{equation}
The second term is well behaved when $c\rightarrow a$ since $K$ is
the elastic response of the coated substrate: the contact edge
singularity results from the boundary conditions, not the response
function. Thus it is $g'$ which is peaked around
$a$~\cite{Barthel98} so that in the end for $c<r<a$
\begin{equation}
\theta'(r)\simeq\frac{2g'(r)}{\Eus}
\end{equation}
and {\it again}
\begin{equation}
  w=\frac{2g(a)^2}{\pi E^\star a} \label{selfconsistenceInhom}
\end{equation}
even for a non-homogeneous substrate.

The local response is indeed dominated by the local stress
distribution and the {\it film} compliance: the same relation
Eq.~\ref{EqSelfConHomo2} between adhesion energy and stress
intensity factor applies.

\subsection{Relation between $g(a)$ and $S$}
Comparing Eqs.~\ref{EqCompliance} and \ref{selfconsistenceInhom},
one infers
\begin{equation}
  \frac{4 g(a)^2}{\Eus}=\frac{1}{2}\delta_{fp}^2 \frac{dS}{da}
\end{equation}
This is a remarkable result which one should in principle be able to
derive from Eq.~\ref{EqEquilibriumFPInhom}.

Note however that the present quasi-analytical
method~\cite{Perriot04} simultaneously provides the flat punch
stiffness and and contact edge stress singularity $g(a)$ (or the
stiffness derivative through Eq.~\ref{EqEquilibriumFPInhom}) by the
simple resolution of a linear system.

\subsection{Normalized Solution}
The procedure to compute a full force curve for the adhesive contact
to coated substrates is therefore: for a given value of the contact
radius $a$
\begin{enumerate}
  \item compute the adhesionless penetration and force for the given indenter
  shape
  \item compute the force and stress intensity factor (or $g(a)$) for the flat punch for a unit value
  of the penetration
  \item compute actual value of $g(a)$ from
  Eq.~\ref{selfconsistenceInhom} and rescale flat punch force and penetration
  \item compute the solution from Eqs.~\ref{EqPenetrationCS} and \ref{EqForceCS}
\end{enumerate}

This provides the solution under the form of two relations between
the local deformation (stress intensity factor) and the local
response on the one hand and the remote loading and the macroscopic
response (contact stiffness) on the other hand.

In normalised form, with
\begin{equation}
\tilde r=\frac{r}{t}
\end{equation}
one has\footnote{For a cone: \begin{eqnarray}
  F_{0,c}&=&\frac{\pi a^2 \Eus}{4\tan\beta}\Pi_c(\tilde a,t,[E])\nonumber\\
  \delta_{0,c}&=&\frac{\pi a}{2\tan\beta}\Delta_c(\tilde
  a,t,[E])\nonumber
\end{eqnarray}
}
\begin{eqnarray}
  F_{0,s}&=&\frac{a^3 \Eus}{2R}\Pi_s(\tilde a,t,[E])\\
  \delta_{0,s}&=&\frac{a^2}{R}\Delta_s(\tilde a,t,[E])\\
  S&=&2 a \Eus \Eeq(\tilde a,t,[E])\\
  g(r;a,t,[E])&=&\frac{\delta_{fp}\Eus}{2}\Gamma(\tilde r;\tilde
  a,t,[E]) \label{Eq_Def_Gamma}
\end{eqnarray}

where the normalized variables can be numerically calculated by the
simple algorithm presented in~\cite{Perriot04}.

All variables equal $1$ for the homogeneous substrate except
$\Pi_s=8/3$ for the sphere\footnote{For a cone $\Pi_c=2$}.

$\Gamma(\tilde r;\tilde a,t,[E])$ is the surface stress transform
normalized to penetration $\delta$. In particular, $\Gamma(\tilde
a;\tilde a,t,[E])$, denoted $\Gamma(1)$ below for brevity, is the
contact edge singularity $g(a;a,t,[E])$ incurred for a coated system
-- normalized to a homogeneous material with the film elastic
properties -- at identical $\delta_{fp}$ value. The variable
$\Gamma(1)$, which is positive  since both $\delta_{fp}$ and $g(a)$
are negative in Eq.~\ref{Eq_Def_Gamma}, is a function of $\tilde a$,
and depends upon the mechanical parameters of the system. From
Eq.~\ref{EqCompliance} the following identity holds:
\begin{equation}\label{EqGde1}
\Gamma(1)
=\sqrt{\frac{1}{2\Eus}\frac{dS}{da}}=\sqrt{\frac{d(a\Eeq)}{da}}
\end{equation}.

Then for the sphere, keeping the Maugis normalization, one
introduces the film thickness normalized to the typical adhesive
contact radius
\begin{equation}
  \eta=\frac{t}{\left(\frac{\pi w R^2}{\Eus}\right)^{1/3}}
\end{equation}
and with $\bar a = \eta\tilde a$ one obtains
\begin{eqnarray}
  \Pi_s &=& \left(\eta\tilde a\right)^3 \frac{\Pi_{s,0}}{2} - 2\sqrt{2} \left(\eta\tilde a\right)^{3/2} \frac{\Eeq}{\Gamma(1)}\label{EqSolNormF} \\
  D_s &=& \left(\eta\tilde a\right)^2 \Delta_{s,0} - \sqrt{2} \left(\eta\tilde a\right)^{1/2} \frac{1}{\Gamma(1)}\label{EqSolNormD}
\end{eqnarray}

These equations generalize
Eqs.~\ref{EqForceJKR}-\ref{EqPenetrationJKR}. The homogeneous
substrate force and penetration terms are corrected by the coated
system factors $\Pi_{s,0}$ and $\Delta_{s,0}$, the homogeneous
contact stiffness by $\Eeq$ and, for identical stress intensity
factor, the homogeneous penetration is corrected by $1/\Gamma(1)$.

\section{Examples of Numerical Results}\label{Results}
\subsection{Reduced variables}
\begin{figure}
\begin{center}
\includegraphics[width=3.25in]{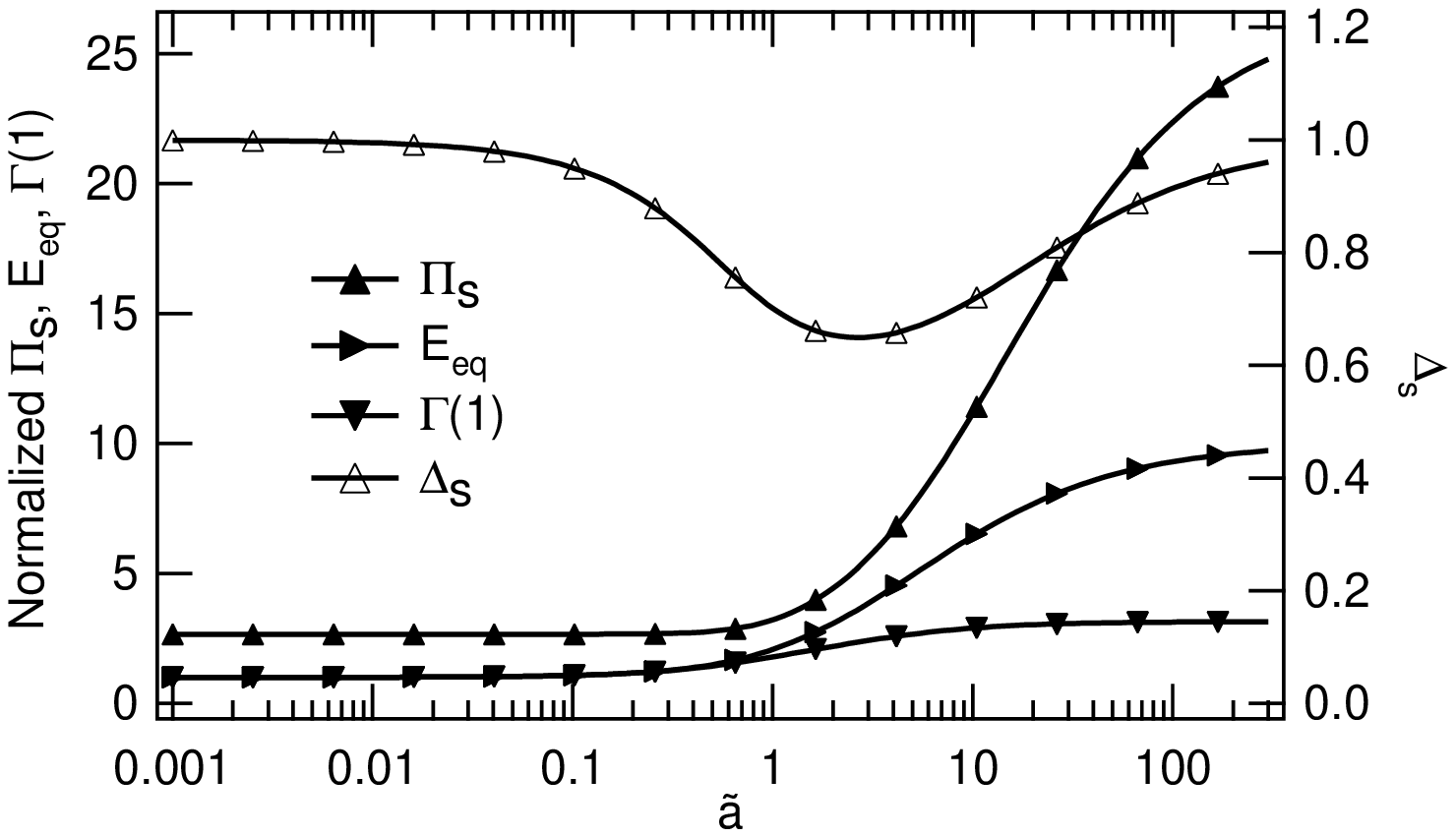}
\caption{}\label{FigTrans10}
\end{center}
\end{figure}

\begin{figure}
\begin{center}
\includegraphics[width=3.25in]{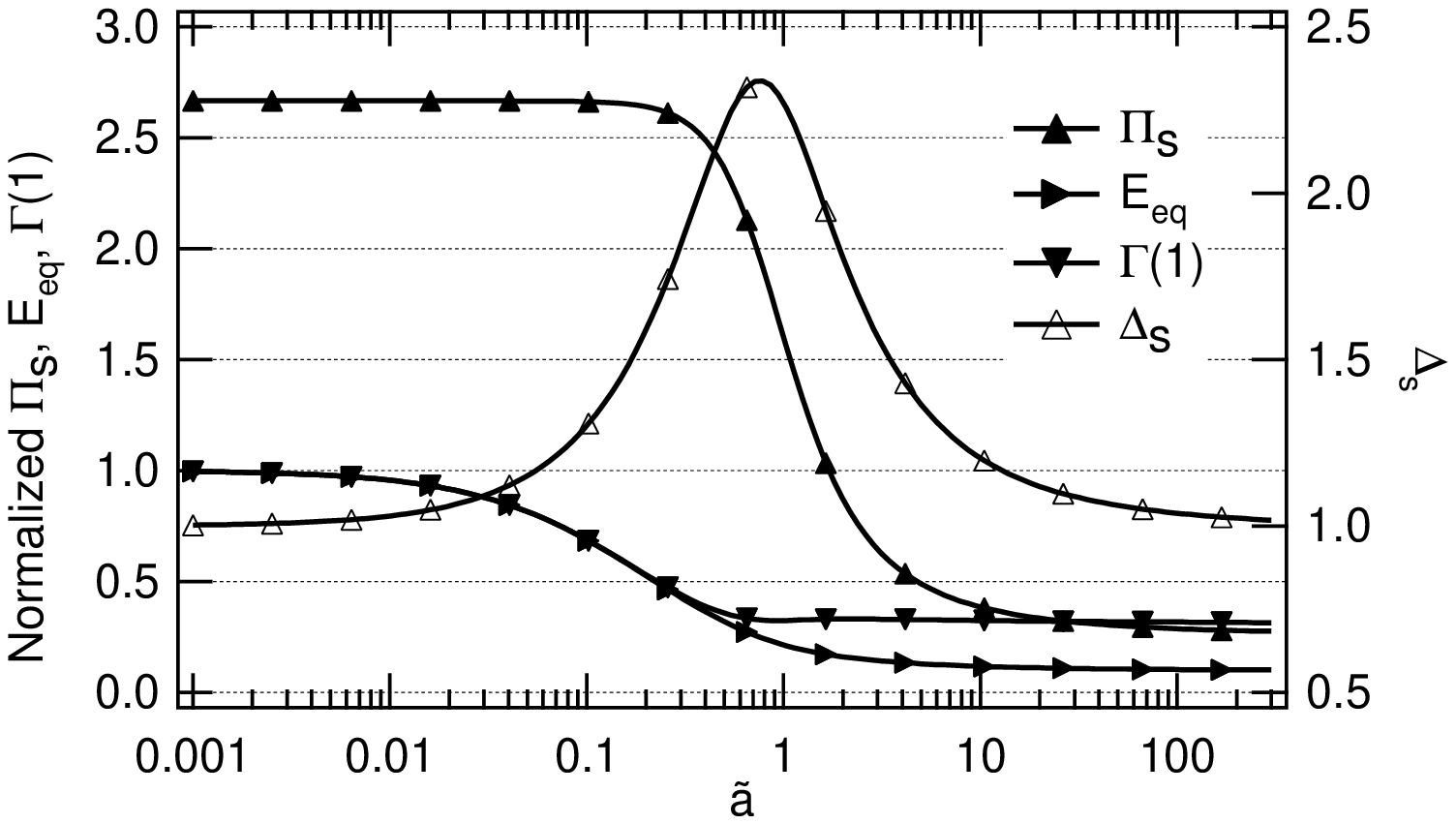}
\caption{}\label{FigTransp1}
\end{center}
\end{figure}

For given mechanical parameters $\Eus/\Ezs$, $\nu_0$ and $\nu_1$ one
may calculate the four variables $\Pi_{s,0}$,$\Delta_{s,0}$,$\Eeq$
and $\Gamma(1)$ as a function of $\tilde a = a/t$. Typical results
are illustrated in Fig.~\ref{FigTrans10} for $\Eus/\Ezs=10$ and
Fig.~\ref{FigTransp1} for $\Eus/\Ezs=0.1$. The results for these
reduced variables compare well with the FE calculations by Sridhar
and Johnson~\cite{Sridhar04}.

The thin film contact problem exhibits a transition between film
dominated to substrate dominated behaviour. In each limit cases, the
contact behaves like a contact to a homogeneous system. The non
trivial behaviour is apparent in the transition which, roughly
speaking, occurs for $\tilde a\simeq 1$, but is shifted to higher
values for compliant films (Fig.~\ref{FigTrans10}) and to lower
values for stiff films (Fig.~\ref{FigTransp1})~\cite{Perriot04}.

$\Eeq$ exhibits a transition between film to substrate reduced
modulus as the contact radius increases is consistent with numerous
previous works~\cite{Mencik97}. The behaviour of $\Delta_{s,0}$,
which deviates from 1 in the midst of the
transition~\cite{Perriot04,Barthel06} has been much less studied, as
mentioned previously (section~\ref{SecIntro}). Similarly for the
flat punch we note that $\Gamma(1)$ tracks $\Eeq$ at small contact
radius values, up to about the middle of the transition. This does
result from Eq.~\ref{EqGde1} as a linear expansion of $\Eeq$ shows
(note that $\Eeq(0)=1$). For larger contact radius values, higher
order terms come into play and $\Gamma(1)$ starts to deviate from
$\Eeq$.

\subsection{Adhesive contact solutions}
The reduced variables $\Pi_s(\tilde a,t,[E])$, $\Delta_s(\tilde
a,t,[E])$, $\Eeq(\tilde a,t,[E])$ and $\Gamma(1)=\Gamma(\tilde
a;\tilde a,t,[E])$ provide the normalized solution
(Eqs.~\ref{EqSolNormF} and \ref{EqSolNormD}) as a function of $\eta$
for a set of mechanical parameters. It is worth emphasizing that --
provided the reduced variables have been calculated on a wide enough
range of values of $\tilde a = a/t$ -- the solution for any values
of contact radius, film thickness and adhesion energy can be
calculated from the reduced variables through the elementary
arithmetics of the normalization Eqs.~\ref{EqSolNormF} and
\ref{EqSolNormD}. This should provide for an easy algorithm to fit
experimental data for adhesive contact on thin films.

\begin{figure}
\begin{center}
\includegraphics[width=3.25in]{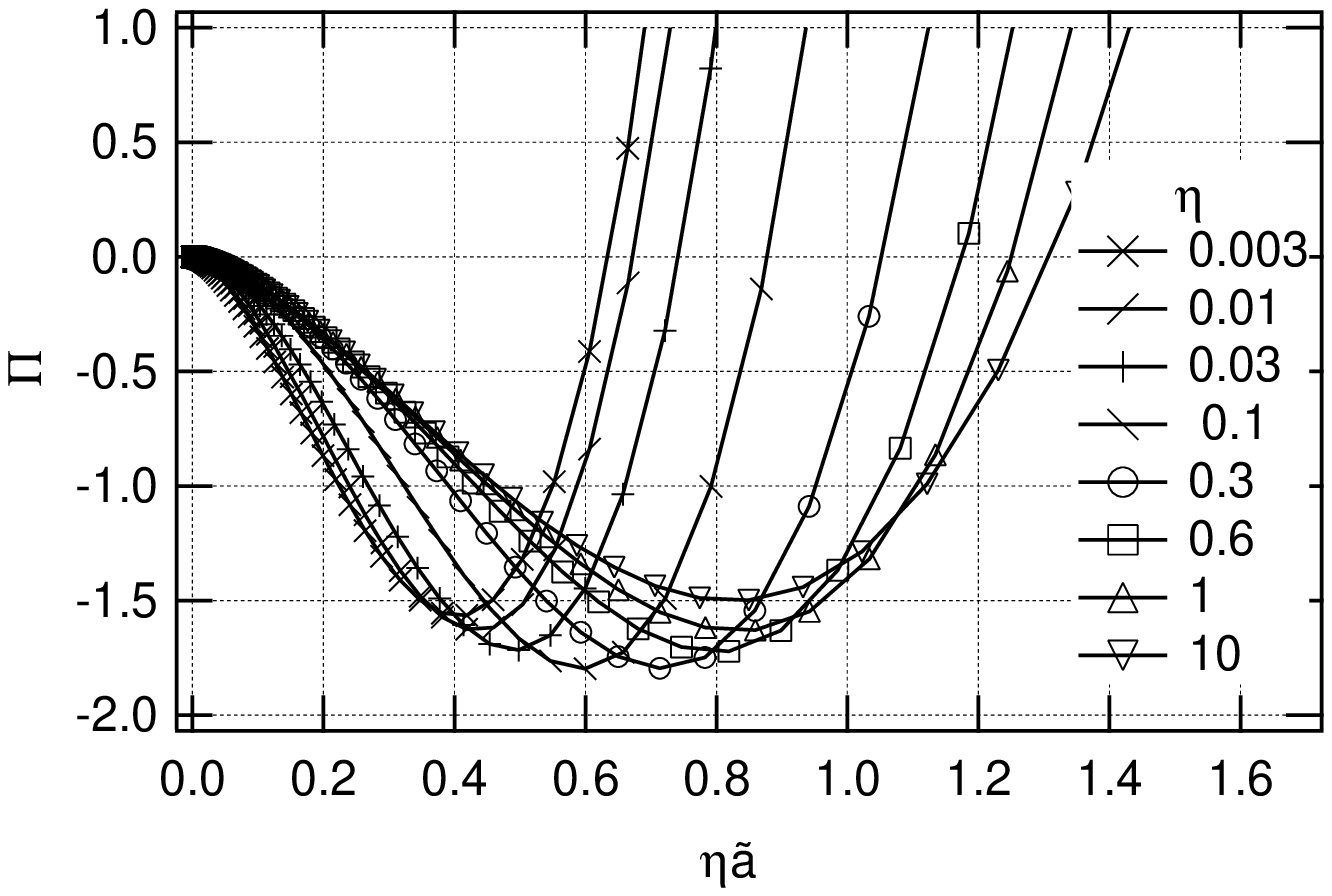}
\caption{}\label{FigAdh10}
\end{center}
\end{figure}

\begin{figure}
\begin{center}
\includegraphics[width=3.25in]{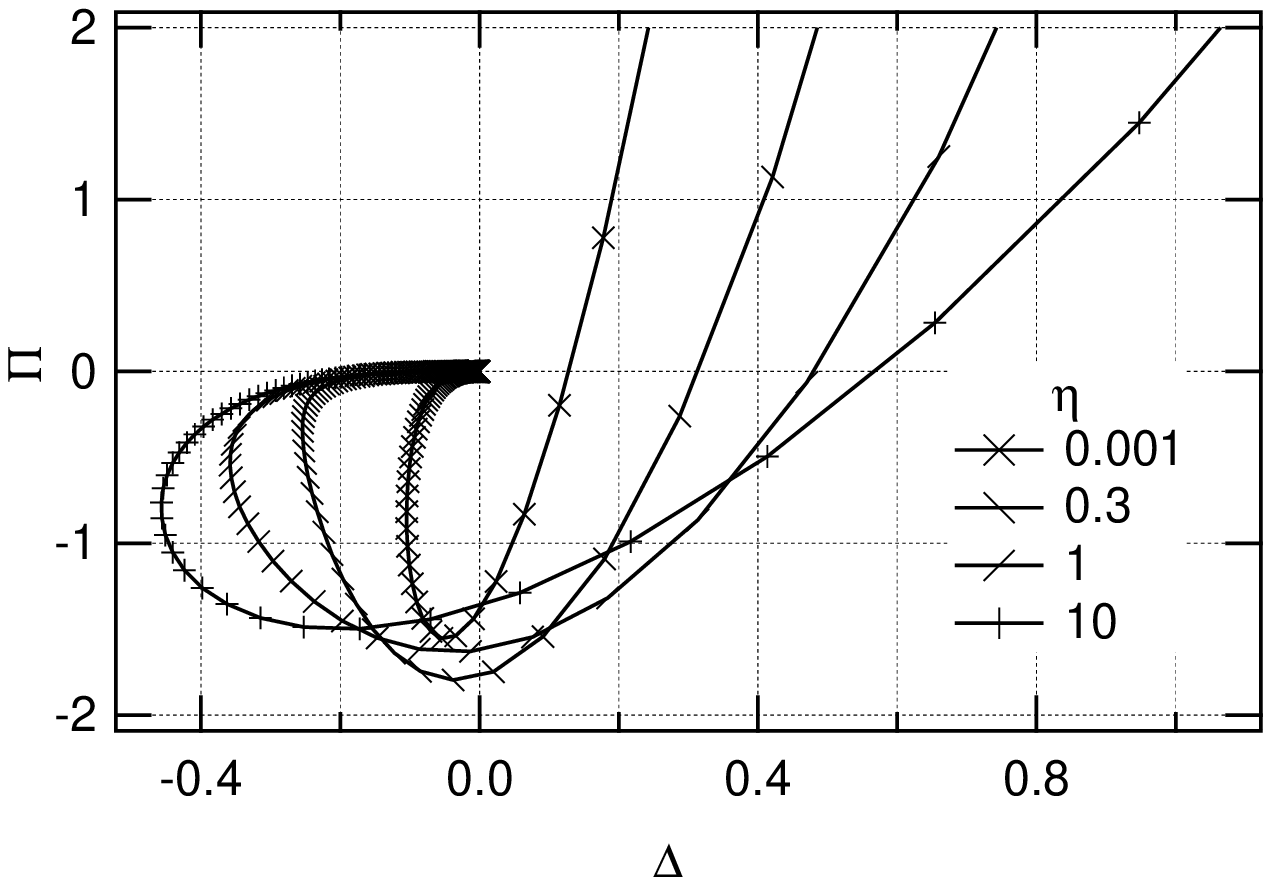}
\caption{}\label{FigPdeD10}
\end{center}
\end{figure}

\begin{figure}
\begin{center}
\includegraphics[width=3.25in]{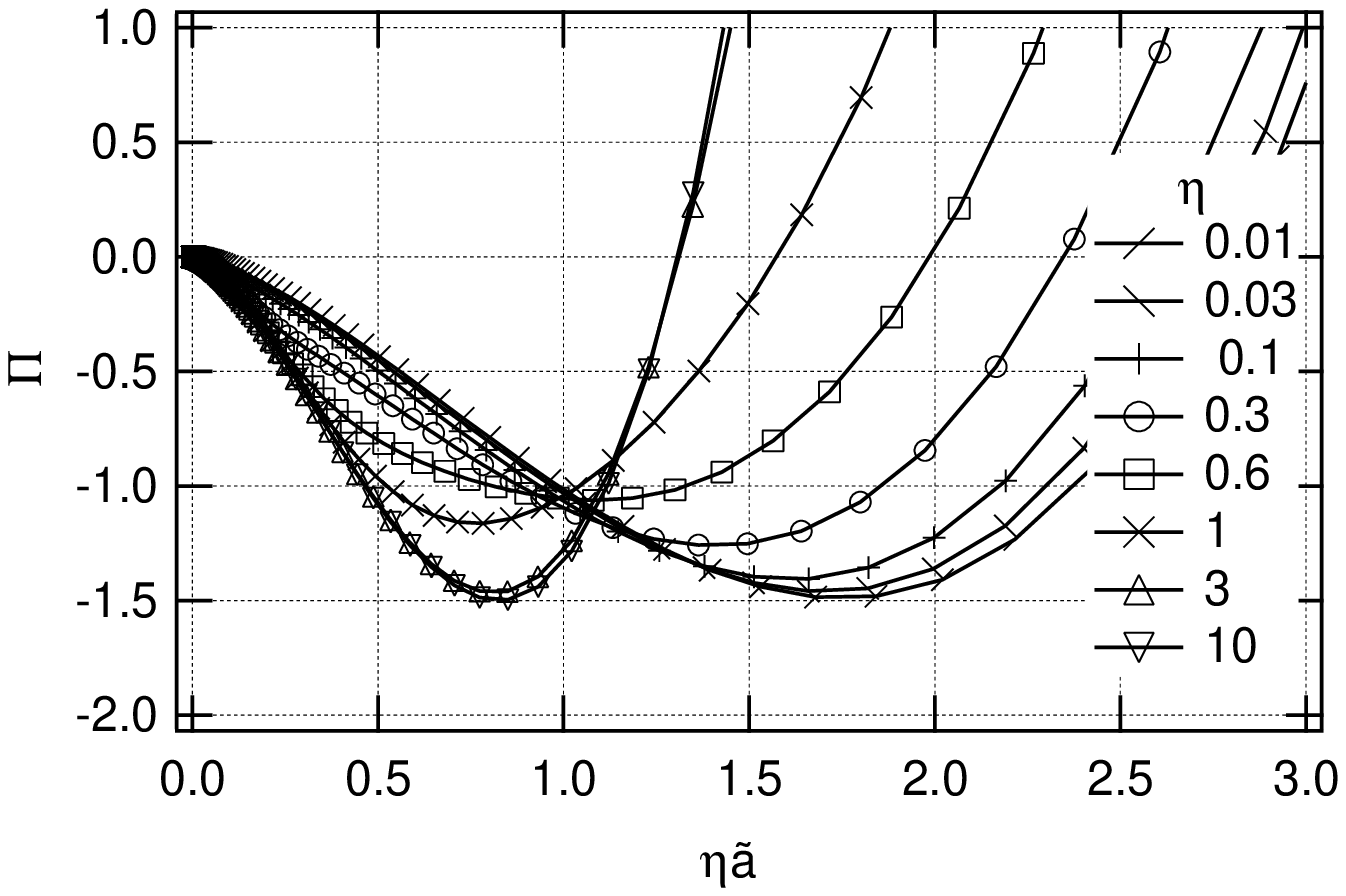}
\caption{}\label{FigAdhp1}
\end{center}
\end{figure}

Results for various values of $\eta$ are displayed on
Fig.~\ref{FigAdh10} and \ref{FigPdeD10} for $\Eus/\Ezs=10$ and
Fig.~\ref{FigAdhp1} for $\Eus/\Ezs=0.1$ with $\nu_0=\nu_1=0.25$.

For a coated substrate in the presence of adhesion, the system
undergoes two transitions. The first one is is the transition
between coating and substrate response~\cite{Perriot04}, when
$\tilde a \simeq 1$ as described in the previous section. The second
one is the transition from adhesion dominated to non adhesive
contact as the load and therefore the contact radius
increase~\cite{Johnson97}. This transition occurs when $\bar a\simeq
1$. The deviation from the JKR results occur when the two
transitions are simultaneous, {\it i.e.} when $\eta \simeq 1$.

Explicitly, for large values of $\eta$ ({\it i.e.} thick films), the
adhesive stage of the contact will occur for small $\tilde a$ and
the contact is film dominated and behaves as a homogeneous material
with the mechanical parameters of the film.

For small values of $\eta$ ({\it i.e.} thick films), the adhesive
stage occurs for large $\tilde a$ and the contact is substrate
dominated: the contact behaves as a homogeneous material with the
mechanical parameters of the substrate. However, at the edge of the
contact, this contact is not equivalent to an adhesive contact to a
homogeneous material with substrate properties. Indeed, the stress
intensity factor is dominated by the film properties
(Eq.~\ref{selfconsistenceInhom}).

As a result, moderate deviations on the pull-out force are
evidenced. For soft layers, the adhesion force is
enhanced~\cite{Sridhar04}. For rigid layers there is a small
reduction of the adhesion force\footnote{This is not, however, the
reason why on soft materials a rigid layer may drastically reduce
the adhesion, which is often due to the suppression of non elastic
additional dissipation phenomena. Note that here the adhesive force
reduction is accompanied by an increase of the stress intensity
factor.}.

\section{Conclusion}

The surface stress transform is adequate to handle the adhesive
contact to coated substrates. It provides a numerically simple
method to compute the four quantities necessary for the actual
description of such adhesive contacts. For a given layer and
substrate mechanical properties, fits to data with free adhesion
energies and coating thicknesses may be performed easily. It also
allows a consistent description of the crack tip which connects the
various concepts developed in adhesive contact problems.

The method also allows the inclusion of finite indenter stiffness, a
question which arises in practice and  which will be studied in more
details in a subsequent paper.

\ack The authors thank A. Chateauminois and C. Fr\'etigny for
several enlightening discussions on this and other related topics.

\section{Appendix: contact stiffness and flat punch stiffness}\label{SecAppendix}
The identity of the contact stiffnesses defined by
Eq.~\ref{EqStiffnessInhom} and by Eq.~\ref{EqRaideur} may look
questionable: is the contact stiffness for an adhesionless curved
indenter identical with the flat punch stiffness for an identical
contact radius, even for an inhomogeneous substrate ?

Since the curved indenter is adhesionless, $g(a)=0$. Then from
Eq.~\ref{EqForceGeneral} we have
\begin{equation}
  \frac{dP}{d\delta}=4\int_0^a\frac{dg(r)}{d\delta}dr
\end{equation}
From the differentiation of Eq.~\ref{EqEquilibriumCS} with $\delta$
together with Eq.~\ref{EqDefTheta},
\begin{equation}
  1=\frac{2}{\pi}\int_0^a\frac{dg(r)}{d\delta}K(r,s;t,[E])dr
\end{equation}
This is the flat punch equilibrium equation for unit penetration.
Then the force
\begin{equation}
  F_{fp}=4\int_0^a\frac{dg(r)}{d\delta}dr
\end{equation}
is the flat punch force for unit penetration, {\it i.e.} the flat
punch stiffness $S(a)$. Therefore for an adhesionless curved
indenter
\begin{equation}
  \frac{dP}{d\delta}=S(a)
\end{equation}
Note however that in general this stiffness is {\em not} the
hertzian $3PR/4a^3$.

\newpage
\bibliographystyle{Lang}
\bibliography{D:/data/Biblio2/Indentation,D:/data/Biblio2/ContactAdhesion,D:/data/Biblio2/ForcesdeSurface,D:/data/Biblio2/mecacouche,D:/data/Biblio2/MecaSolGel,D:/data/Biblio2/Materiaux,D:/data/Biblio2/Mecanique}

\begin{thebibliography}{10}

\bibitem{Perriot04}
Perriot A. and Barthel E.{\em J. Mat. Res.}, {\bf 2004}, {\em 19}, 600--608.

\bibitem{Saha02}
Saha R. and Nix W.~D.{\em Acta Materialia}, {\bf 2002}, {\em 50}, 23--38.

\bibitem{Jung04}
Jung Y.-G., Lawn B.~R., Martyniuk M., Huang H., and Hu X.~Z.{\em J. Mater.
  Res.}, {\bf 2004}, {\em 19}, 3076.

\bibitem{Han06}
Han S.~M., Saha R., and Nix W.~D.{\em Acta Materialia}, {\bf 2006}, {\em 54},
  1571--1581.

\bibitem{Barthel06}
Barthel E., Perriot A., Chateauminois A., and Fr{\'e}tigny C.{\em Phil. Mag.}, {\bf 2006},
  {\em 86}, 5359-5369.

\bibitem{Derjaguin34}
Derjaguin B.{\em Kolloid-Zeitschrift}, {\bf 1934}, {\em 69}, 55--164.

\bibitem{JKR}
Johnson K.~L., Kendall K., and Roberts A.~D.{\em Proc. Roy. Soc. London A},
  {\bf 1971}, {\em 324}, 301.

\bibitem{DMT}
Derjaguin B.~V., Muller V.~M., and Toporov Yu.~P.{\em J. Colloid Interface
  Sci.}, {\bf 1975}, {\em 53}, 314.

\bibitem{Maugis00}
Maugis D., {\em Contact, Adhesion and Rupture of Elastic Solids} (Springer,
  Berlin Heidelberg, 2000).

\bibitem{Johnson01}
Johnson K.~L. and Sridhar I.{\em J. Phys. D: Appl. Phys.}, {\bf 2001}, {\em
  34}, 683--689.

\bibitem{Sridhar04}
Sridhar I., Zheng Z.~W., and Johnson K.~L.{\em J. Phys. D: Appl. Phys.}, {\bf
  2004}, {\em 37}, 2886--2895.

\bibitem{Mary07}
Mary P., Chateauminois A., and Fr{\'e}tigny C.{\em J. Phys. D: Appl. Phys.},
  {\bf 2006}, {\em 39}, 3665.

\bibitem{Sperling64}
G.~Sperling.
\newblock {\em Eine Theorie der Haftung von Feststoffteilchen an festen
  Koerpern}.
\newblock PhD thesis, T.U. Karlsruhe, 1964.

\bibitem{Johnson85}
Johnson K.L., {\em Contact Mechanics} (Cambridge University Press, Cambridge,
  1985).

\bibitem{Irwin57}
Irwin G.~R.{\em J. Appl. Mech.}, {\bf 1957}, {\em 24}, 361.

\bibitem{Maugis78}
Maugis D. and Barquins M.{\em J. Phys. D.: Appl. Phys.}, {\bf 1978}, {\em 11},
  1989.

\bibitem{Maugis87}
Maugis D.{\em J. Adhesion Sci. Tec.}, {\bf 1987}, {\em 1}, 105.

\bibitem{Greenwood81}
Greenwood J.~A. and Johnson K.~L.{\em Phil. Mag.}, {\bf 1981}, {\em 43}, 697.

\bibitem{Maugis92}
Maugis D.{\em J. Colloid Interface Sci}, {\bf 1992}, {\em 150}, 243.

\bibitem{Barthel02}
Barthel E. and Haiat G.{\em Langmuir}, {\bf 2002}, {\em 18}, 9362--9370.

\bibitem{Haiat03}
Haiat G., Huy M.~C.~Phan, and Barthel E.{\em J. Mech. Phys. Sol.}, {\bf 2003},
  {\em 51}, 69--99.

\bibitem{Ting66}
Ting T.~C.~T{\em J. Appl. Mech}, {\bf 1966}, {\em 33}, 845.

\bibitem{Sergici06}
Onur~Sergici A., Adams G.~G, and Muftu S.{\em J. Mech. Phys. Sol.}, {\bf 2006},
  {\em 54}, 1843--1861.

\bibitem{Shull98}
Shull K.~R., Ahn D. , Chen W.~L., Flanigan C.~M., and Crosby A.~J.{\em
  Macromol. Chem. Phys.}, {\bf 1998}, {\em 199}, 489--511.

\bibitem{Shull02}
Shull Kenneth~R.{\em Mat. Sci. Eng. R: Reports}, {\bf 2002}, {\em 36}, 1--45.

\bibitem{Parks74}
Parks D.~M.{\em Int. J. Fract.}, {\bf 1974}, {\em 10}, 487--502.

\bibitem{Gladwell80}
Gladwell G.~M.~L., {\em Contact Problems in the Classical Theory of Elasticity} (Sijthoff \& Noordhoff, Germantown, 1980).

\bibitem{Barthel98}
Barthel E.{\em Thin Solid Films}, {\bf 1998}, {\em 330}, 27--33.

\bibitem{Huguet00}
Huguet A.~S. and Barthel E.{\em J. Adhesion}, {\bf 2000}, {\em 74}, 143--175.

\bibitem{Greenwood98}
Greenwood J.~A. and Johnson K.~L.{\em J. Phys. D: Appl. Phys.}, {\bf 1998},
  {\em 31}, 3279.

\bibitem{Barthel04}
Barthel E. and Haiat G.{\em J. Adhesion}, {\bf 2004}, {\em 80}, 1.

\bibitem{Schapery89}
Schapery R.~A.{\em Int. J. Fract.}, {\bf 1989}, {\em 39}, 163.

\bibitem{Mencik97}
Mencik J., Munz D., Quandt E., Weppelmann E.~R., and Swain M.~V.{\em J. Mater.
  Res.}, {\bf 1997}, {\em 12}, 2475--2484.

\bibitem{Johnson97}
Johnson K.~L. and Greenwood J.~A.{\em J. Colloid Interface Sci.}, {\bf 1997},
  {\em 192}, 326--333.

\end{thebibliography}
\newpage
Captions

Fig.~1: schematics of an adhesive contact: the contact zone (radius
$a$) is surrounded by the interaction zone. The gap between surfaces
outside of the contact zone is $h(r)$. The normal stress
distribution turns from compressive at the center of the contact to
tensile on both sides of the contact edge. The transform
$\theta(r_1)$ is calculated from the normal surface displacement at
$r<r_1$. The transform $g(r_2)$ is calculated from the normal
surface stress for $r>r_2$.

Fig.~2: Adhesive contact configuration obtained from the
adhesionless contact by a flat punch displacement $\delta_{fp}$
along the tangent to the adhesionless force~$F$-penetration~$\delta$
curve. Integration gives the stored elastic energy. The energy
release rate is obtained as the variation with contact radius $a$ of
the stored elastic energy at constant total penetration $\delta$
(shaded area).

Fig.~3: Normalized force $\Pi_s$ (left) and penetration $\Delta_s$
(right) for an adhesionless sphere as a function of contact radius
normalized to film thickness $\tilde a=a/t$. The film to substrate
modulus ratio is 0.1 and Poisson ratio 0.25 for both materials. The
contact stiffness $E_{eq}$ (left) and contact edge stress intensity
factor $\Gamma(1)$ (left) are calculated from the flat punch
boundary conditions. For $\tilde a \simeq 1$ the transition from
film dominated to substrate dominated contact occurs.

Fig.~4: Same plot as Fig.~3 for a film to substrate modulus ratio
equal to 10. The transition occurs earlier due to film rigidity.

Fig.~5: Normalized force versus normalized contact radius as a
function of $\eta$, the ratio of the film thickness to the typical
zero load adhesive contact radius. When $\eta\simeq 1$, adhesive
effects are significant in the transition from film to substrate
dominated regimes: deviations from the JKR model are observed and
increased adhesion forces are observed.

Fig.~6: Normalized force versus normalized penetration as a function
of $\eta$. Identical parameters as in Fig.~5.

Fig.~7: Similar plot as in Fig.~5 with a film to substrate modulus
ratio of 10. In the intermediate regime $\eta\simeq 1$, adhesion
forces are reduced.

\newpage

\end{document}